\documentclass[twocolumn,amsmath,amssymb]{revtex4}
\usepackage{graphicx}
\usepackage{dcolumn}
\begin{document}
\title{Teleportation in the presence of common bath decoherence at the transmitting station}
\author{D. D. Bhaktavatsala Rao, P. K. Panigrahi and Chiranjib Mitra}
\affiliation{Indian Institute of Science Education and Research Kolkata, INDIA.}
\date{\today}
\begin{abstract}
We investigate the effect of common bath decoherence on the qubits of Alice in the usual teleportation protocol. The system bath interaction is studied under the central spin model, where the qubits are coupled to the bath spins through isotropic Heisenberg interaction. We have given a more generalized representation of the protocol in terms of density  matrices and calculated the average fidelity of the teleported state for different Bell state measurements performed by Alice. 
The common bath interaction differentiates the outcome of various Bell state measurements made by Alice. There will be a high fidelity teleportation for a singlet measurement made by Alice when both the qubits of Alice interact either ferromagnetically or antiferromagnetically with bath. In contrast if one of the Alice's qubits interact ferromagnetically and the other anti-ferromagnetically then measurement of Bell states belonging to the triplet sector will give better fidelity.  We have also evaluated the average fidelity when Alice prefers non-maximally entangled states as her basis for measurement.
\end{abstract} 
\pacs{Valid PACS appear here}
\maketitle
\section{Introduction}
Quantum protocols are mostly designed in the idealistic situation of a decoherence free system. In practical implementation of these protocols the external environment can play a significant role in reducing the fidelity of the expected outcomes. The various environmental interactions of the quantum system are dealt using either a harmonic oscillator bath or a spin bath \cite{book}. 
At low enough temperatures where only a few energy modes of environmental constituents contribute to the dynamics, modeling the bath by a set of spins (finite dimensional) seems to be a more natural choice \cite{spbath}. On the other hand it was shown that spin environments always lead to non-Markovian decoherence of the quantum system,  where the decoherence time scales and saturation value of qubit polarization will depend on the initial polarizations of the both the system and bath spins \cite{durga1, bruer}. This dependence can have considerable effect on quantum protocols, we shall discuss one such situation later in this paper.

In implementing quantum teleportation \cite{bennet}, Bob gets the same unknown state after a proper unitary transformation irrespective of the Bell state measurement made by Alice. The same may not be true in presence of environmental interaction both at the transmitting and the receiving station. Most of the recent work has been devoted in studying the effect of local noises for both the Alice and Bob's qubit \cite{recent1, recent2}.
It is important to note that Alice is in possession of two qubits where as Bob has only one. Hence the decoherence at the transmitting station can be quite different from that of the receiving station. If the qubits of Alice see different environments, there is nothing new to the dynamics as Bob gets the same decohered state irrespective of the Bell state shared with Alice. On the other hand if both the qubits of Alice see a common environment the situation will be quite different, as there will be bath mediated interaction between the two. This can lead to the bath induced entanglement between Alice's qubits. As the common environment for Alice's qubits seems to be a more natural choice to study the effects of decoherence on teleportation, the present study aim at explicating such effects through an exactly solvable model described below. The problem of common bath decoherence becomes more relevant in cases where Alice and Bob share $n$ qubit entangled state with $n-k$ qubits at Alice possession and $k$ with Bob \cite{nent}. The present work can be a starting point for all such generalized studies 

\subsection{The Model}
In this paper we shall consider the central spin model to study the effect of spin bath on the qubits used for teleportation. This model can be more realistic in quantum dot systems where the dominant contribution to decoherence comes from the nuclear spins interacting through the homogeneous Heisenberg interaction with the system spins. Recently it was shown how quantum dots can be used as a resource for teleportation \cite{qdottele1, qdottele2}. The present study can also be relevant for other solid state systems using spins as qubits.

The Hamiltonian describing the interaction between the qubits and the spin bath is given by
\begin{eqnarray}
\label{ham}
H = K_a \vec{S}_a \cdot \vec{I}_{\mathcal{E}a} + K_A \vec{S}_A \cdot \vec{I}_{\mathcal{E}A} + K_B \vec{S}_B\cdot \vec{I}_{\mathcal{E}B},
\end{eqnarray}
where $\vec{S}_a, \vec{S}_A$, represents the spin operators of the two qubits at the transmitting site which are in possession of Alice and $\vec{S}_B$ the spin operator of Bob's qubit. The total spin of the environmental particles seen by each qubit is represented by $\vec{I}_{\mathcal{E}} = \sum_k\vec{I}_{\mathcal{E},k}$. The interaction strengths of the qubits with their respective baths are denoted by $K_a$, $K_A$ and $K_B$. Depending on the common set of environmental spins seen by Alice's qubits, one has either a common spin bath or separate baths. The decoherence at Bob's site is trivial since Bob has only qubit, we set $K_B=0$. The interest is on the transmitting side as there are two qubits. It is well known from the earlier studies that a common bath can induce entanglement between two initially unentangled qubits \cite{benati, durga2}. The common bath interaction between the qubits can lead to a natural selection of two-qubit states which are less prone to decoherence \cite{durga2}. Hence all these effects can show considerable effect on the fidelity of the transmitted state.

In this work we shall restrict ourselves completely to the density matrix notation. Writing the qubit states explicitly in terms of the polarizations and correlations helps us understand the underlying physical structure in a straight forward manner.
The most general representation of a two-qubit Bell state is given by
\begin{eqnarray}
\rho_{AB}=\frac{1}{4}\hat{\mathcal{I}} + \sum_k D_k S^k_AS^k_B,
\end{eqnarray}
where $S^k_A$ and $S^k_B$ are the spin operators of the qubits $A$ and $B$ respectively. The correlation vector $D_k \equiv {\rm Tr}(\rho_{AB}S^k_AS^k_B)$. For the Bell states the corresponding correlation vectors are given by $\vec{D}_{S_0} = [-1,-1,-1]$, $\vec{D}_{T_0} = [1, 1, -1]$, $\vec{D}_{T_+} = [1, -1, 1]$, and $\vec{D}_{T_-} = [-1, 1, 1]$. The Bell states are represented in the $\hat{z}$ basis of the two qubits as $|S_0\rangle =\frac{1}{\sqrt{2}}[|\uparrow\downarrow-\downarrow\uparrow\rangle]$, $|T_0\rangle = \frac{1}{\sqrt{2}}[|\uparrow\downarrow+\downarrow\uparrow\rangle]$, 
$|T_+\rangle =\frac{1}{\sqrt{2}}[|\uparrow\uparrow+\downarrow\downarrow\rangle]$ and 
$|T_-\rangle = \frac{1}{\sqrt{2}}[|\uparrow\uparrow-\downarrow\downarrow\rangle]$. It can be immediately seen that one of them belongs to the singlet sector and the other three to the triplet sector. When evolving through Hamiltonian dynamics each of them respond differently to the environmental interaction and to the applied fields.

\subsection{Decoherence free Teleportation}
Using the Bell-state representation given above the initial state used in the teleportation protocol is given by
\begin{eqnarray}
\label{istate}
 \rho_{aAB} &=& \rho_a \otimes \rho_{AB} \nonumber \\
&=&\frac{1}{2}\left[\hat{\mathcal{I}}+2\vec{P}_a\cdot\vec{S}_a\right]\otimes \frac{1}{4}\left[\hat{\mathcal{I}} + 4\sum_k D_k S^k_AS^k_B \right],
\end{eqnarray}
where $\vec{P}_a = {\rm Tr}\rho_a\vec{S_a}$ is the polarization vector of the unknown state which is to be teleported to Bob.
Now Alice performs a Bell state measurement on her two qubits $a$ and $A$, i.e.,
\begin{eqnarray}
{\rm Tr}_{aA}[\rho_{aA}\otimes\hat{\mathcal{I}}_B \rho_{aAB}],
\end{eqnarray}
where $\rho_{aA} = \frac{1}{4}\hat{\mathcal{I}} + \sum_k M_k S^k_aS^k_A$ is the Bell state of qubits $a$, $A$.
The vector $\vec{M}$ has the information of the Bell measurement made by Alice. After performing the trace the state that Bob gets with $1/4$ probability is given by
\begin{eqnarray}
 \rho_B = \frac{1}{2}\left[\hat{\mathcal{I}}+2\vec{P}_B\cdot\vec{S}_B\right].
\end{eqnarray}
The polarization of Bob's qubit is related to the polarization of the unknown state through the correlation vectors $D$ and $M$
as
\begin{eqnarray}
\label{dfbpol}
 P^i_B = D_iM_iP^i_a.
\end{eqnarray}
Now if the vectors $\vec{D}$ and $\vec{M}$ are equal, Bob needs to do nothing to this qubit. If $(\vec{D}\times\vec{M})\cdot\hat{n} \ne 0$, then Bob has to do a rotation of his qubit along $\hat{n}$ direction. Performing the appropriate unitary transformation is  equivalent to multiplying $D_iM_i$ to $P^i_B$ in Eq.\ref{dfbpol}. 
Since $D^2_i = M^2_i = 1$ we immediately see that the final state that of Bob is the unknown state that Alice wishes to teleport i.e., $\rho_B = \rho_a$.

Thus we have given a more general description of the teleportation scheme where Bob's operation has a physical meaning in terms of the plane chosen by the correlation vectors of the $AB$ and $aA$ entangled states. We shall show later that the above representation helps in dealing with the time-evolution of the qubits in the presence of environment more easily.

\section{Common bath decoherence for Alice's qubits}
In this section we shall study the effects of a common bath decoherence for Alice's qubits on fidelity of the teleported state.
The Hamiltonian given in Eq.\ref{ham} reduces to
\begin{eqnarray}
\label{hamsolv}
H = (K_a \vec{S}_a + K_A \vec{S}_A)\cdot\vec{I}_\mathcal{E},
\end{eqnarray}
where $\vec{I}_\mathcal{E}$ represents the total bath spin. 
Note that eventhough there is no direct interaction between Alice's qubits, their interaction with the bath can result in an indirect coupling between the two. This will lead to the generation of entanglement between Alice's qubits prior to her measurement. 

We shall take the initial state of the bath as an incoherent superposition of states labelled by the bath spin $I_{\mathcal{E}}$, with weights $\lambda_{I_{\mathcal{E}}}$, $\rho_{\mathcal{E}}(0)= \sum \lambda_{I_{\mathcal{E}}} \rho_{I_{\mathcal{E}}}(0)$. In this study all $\rho_{I_{\mathcal{E}}}(0)$ will be taken to be unpolarized (multiple of identity). The weights $\lambda_{I_{\mathcal{E}}}$ are however, free parameters.

For implementing the teleportation protocol in the presence of decoherence we shall first rewrite the initial state of the Alice-Bob system given in Eq.\ref{istate} as follows
\begin{eqnarray}
\rho_{aAB}(0) = \frac{1}{8}\hat{\mathcal{I}} + {1\over 2}\vec{P}_a\cdot\vec{S}_a + {1\over 2}{\vec{\mathbb P}}_A\cdot\vec{S}_A + 
\sum_{m,n=1}^{3}{\bf\mathbb{D}}_{mn}S^m_{a}S^n_{A}.
\end{eqnarray}
In writing the above we have absorbed Bob's spin $\vec{S}_B$ into the polarization vectors. The components of the new polarization vectors ${\mathbb P}^i_A=D_iS^i_B$, and $\mathcal{\mathbb D}^{mn} = P^m_aD^nS^n_B$. Since the time-evolution is only for qubits $a$ and $A$, the above form is valid.

The state of the total system before Alice makes the Bell measurement is time-dependent, given by
\begin{eqnarray}
\rho_{aAB}(t) = {\rm Tr}_\mathcal{E}\left(U_{H}(t) \rho_{aAB}(0)\otimes\rho_\mathcal{E} U^\dagger_{H}(t)\right),
\end{eqnarray}
where ${\rm Tr}_\mathcal{E}$ represents the summing over the bath degrees of freedom. Note that the initial state of the system-bath is uncorrelated. In the above equation the unitary operator $U_H$ corresponding to the Hamiltonian in Eq. \ref{hamsolv} is given by
 \begin{eqnarray}
\label{unitop}
U_H&=& \left[ a_1(t)+a_2(t)(\vec{S}_a-\vec{S}_A)\cdot \vec{I}_\mathcal{E} \right] (1-\frac{\hat{S}^2_{aA}}{2}) \nonumber \\
&& +\left [ a_3(t) + a_4(t)\vec{S}_{aA}\cdot \vec{I}_\mathcal{E} + a_5(t)(\vec{S}_{aA}\cdot \vec{I}_\mathcal{E})^2 \right. \nonumber \\
&& + \left. a_6(t)(\vec{S}_a-\vec{S}_A)\cdot \vec{I}_\mathcal{E} 
+ a_7(t)(\vec{S}_a \times \vec{S}_A)\cdot \vec{I}_\mathcal{E} \right ]  \frac{\hat{S}^2_{aA}}{2}. \nonumber \\
\end{eqnarray}
where $\vec{S}_{aA}=\vec{S}_a + \vec{S}_A$. The time-dependent coefficients $a_i(t)$ have been derived earlier in \cite{durga2}. 
After performing the trace over bath degrees of freedom we obtain the reduced density matrix of the Alice-Bob system, given by
\begin{eqnarray}
\hspace{-10mm}
\rho_{aAB}(t) = \frac{1}{8}\hat{\mathcal{I}} + {1\over 2}\vec{P}_a(t)\cdot\vec{S}_a + {1\over 2}{\vec{\mathbb P}}_A(t)\cdot\vec{S}_A + \sum_{m,n=1}^{3}{\bf\mathbb{D}}_{mn}(t)S^m_{a}S^n_{A}. \nonumber
\end{eqnarray}

After Alice makes the Bell measurement on her qubits, the state that Bob gets with $1/4$ probability is given by
\begin{eqnarray}
\rho_B(t) = \frac{1}{2}\hat{\mathcal{I}} + \sum_k M_k \mathbb{D}_{kk}(t).
\end{eqnarray}
The time-dependent coefficient $\mathbb{D}_{kk}(t)$ is given by 
\begin{eqnarray}
\mathbb{D}_{kk}(t) &=& f(t) \mathbb{D}_{kk}(0) + g(t){\rm Tr}[\mathbb{D}(t)], \nonumber \\
&=&f(t) P^k_aD_kS^k_B + g(t)\sum_m P^m_a D_m S^m_B.
\end{eqnarray}
Depending on the two bits of classical information given by Alice, Bob makes the appropriate unitary transformation. The final state of Bob is then given by
\begin{eqnarray}
\rho_B(t)= \frac{1}{2}\hat{\mathcal{I}} + \vec{P}_B(t)\cdot\vec{S}_B,
\end{eqnarray}
where $P^i_B = f(t)P^i_a + g(t)M_i{\rm Tr}M P^i_a$. If system bath interaction is zero i.e., $K_a = K_A =0$ then $f(t)=1$, $g(t)=0$, and we get perfect teleportation. In the presence of the bath Bob's final state depends on $\vec{M}$ from which he can know about the measurement made by Alice. Note that there is no information of $M$ in decoherence free teleportation after Bob has made his final transformation. One can show that if the qubits of Alice see separate environments there will be no such dependence of Alice measurement. Hence common bath has introduced a new feature to the protocol where Bob's final state has the information of the Bell measurement made by Alice.
\begin{figure}[htb]
\begin{center}
   \includegraphics[width=8.0cm]{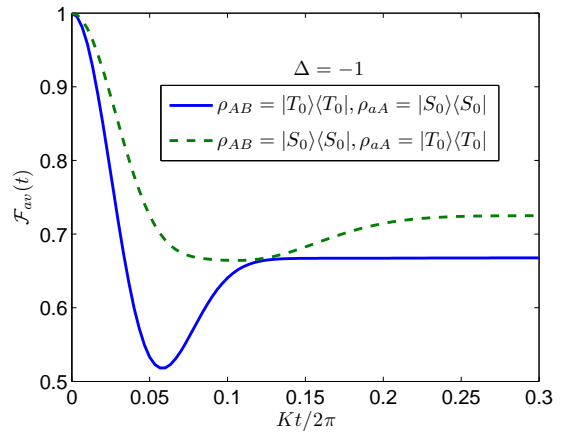}
\end{center}
 \caption{(Color online)We have plotted average fidelity of teleporting the unknown state given to Alice as function of time. We have considered two cases, when the Bell state shared by Alice and Bob belongs to the triplet sector and measurement made by Alice is in the singlet sector and the converse. One of the Alice's qubits interact ferromagnetically and the other anti-ferromagnetically. The interacting strengths are chosen such that $\Delta = -1$. Here $S_0$ and $T_0$ are the singlet and triplet $S^z=0$ states. The bath is completely unpolarized consisting of $N=22$ spins and $K^2 = K^2_a + K^2_A$.}
 \end{figure}
\subsection{Average Fidelity of Teleportation}
In this section we shall calculate the fidelity and average fidelity of teleportation in the presence of decoherence.
Fidelity gives the information of how close is the teleported state to the unknown state, defined as
\begin{eqnarray}
\mathcal{F}(t) &\equiv& \frac{1}{2}\left[1+\vec{P}_a\cdot\vec{P}_B(t)\right], \nonumber \\
&=&\frac{1}{2}\left[1+f(t)|\vec{P}_a|^2 + g(t){\rm Tr}M\sum_k (P^k_a)^2M_k\right].
\end{eqnarray}
By performing an average over all pure states of qubit $a$ we get the average fidelity, which is given by
\begin{eqnarray}
\label{avfid}
\mathcal{F}_{av}(t) &=& \frac{1}{4\pi}\int^{2\pi}_0 d\phi \int^\pi_0 d\theta \sin \theta  \mathcal{F}(t), \nonumber \\
&=& \frac{1}{2}\left[1+f(t) + \frac{1}{3}g(t)({\rm Tr}M)^2\right].
\end{eqnarray}
Though the complete analytical forms for the the time-dependent coefficients $f(t), g(t)$ can be derived exactly we do not give their expressions as they have a  complicated form \cite{durga2}. Instead we restrict to the leading order time-dependence of these coefficients in studying the short time behavior, and for long time behavior we give the appropriate numerical plots.

The leading order time dependence of the coefficients $f(t)$ and $g(t)$ are given by
\begin{eqnarray}
&&f(t) \approx 1-\frac{1}{3}\lbrace \langle \hat{I}^2_{\mathcal{E}} \rangle (K^2_a+K^2_A + K_a K_A ) \rbrace t^2, \nonumber \\
&&g(t) \approx \frac{1}{3}\langle \hat{I}^2_{\mathcal{E}} \rangle K_a K_A t^2.
\end{eqnarray}
If $K_a = K_A$ then $f(t)+3g(t)=1$. Hence for identical couplings as we shall later show that singlet measurement of Alice's qubits always gives perfect teleportation of the unknown state.
Now substituting these time-dependent coefficients in Eq.\ref{avfid}, we find
\begin{eqnarray}
\mathcal{F}_{av}(t) = 1-\frac{t^2}{12}\langle \hat{I}^2_{\mathcal{E}} \rangle(K^2_a+K^2_A)  \left[1 + \Delta(1-({\rm Tr}M)^2/3\right] 
\end{eqnarray}
where the inhomogeneity parameter $\Delta =2K_AK_a/(K^2_A+K^2_a)$ and $\langle \hat{I}^2_{\mathcal{E}}\rangle = \sum_{I_{\mathcal{E}}}\lambda_{I_\mathcal{E}} I_\mathcal{E}(I_\mathcal{E}+1)$. For completely unpolarized baths i.e., $\rho_\mathcal{E} = \frac{1}{2^N}\hat{\mathcal{I}}$, $ \langle \hat{I}^2_{\mathcal{E}} \rangle = 3N/4$, where $N$ is number of bath spins.

For Bell states ${\rm Tr}M$ has only two values, $-3, 1$. If the state is singlet then ${\rm Tr}M = -3$ and for the remaining Bell states it has the value one. The average fidelity has an initial Gaussian decay, $\mathcal{F}_{av}(t) = \exp (-t^2/\tau^2)$, with two different decoherence time scales depending on the Bell state measurement made by Alice, given by
\begin{eqnarray}
\left(\frac{1}{\tau^2}\right)_{S_0} &=& \frac{1}{6}\langle \hat{I}^2_{\mathcal{E}}\rangle( K^2_a+K^2_A)(1-\Delta),\nonumber \\
\left(\frac{1}{\tau^2}\right)_{T_0, T_+, T_-} &=&\frac{1}{6} \langle \hat{I}^2_{\mathcal{E}}\rangle (K^2_a+K^2_A)\left(1+\frac{\Delta}{3}\right).
\end{eqnarray}
The Bell states $S_0, {T_0, T_+, T_-}$ are defined earlier in Sec-IA. When the couplings are identical i.e., $K_a = K_A$, $\Delta =1$ and hence the singlet measurement on Alice's qubits does not harm the state teleported to Bob. On the other hand if the sign of $\Delta$ becomes negative i.e., if one of Alice's qubit is interacting ferromagnetically with the bath and the other antiferromagnetically, then singlet measurement would give a highly decohered state to Bob in comparison to the other Bell measurements. Thus the sign of the interaction with the bath can decide which particular measurement of Alice can give Bob a less decohered state. In Fig.1 we have plotted average fidelity as a function of time for $\Delta = -1$. Contrast to the case of $\Delta=1$, where singlet measurements are preferred to triplet measurements, $\Delta=-1$ prefers triplet measurement to singlet measurement. Thus depending on the nature of interaction of Alice's qubits with the bath, various measurements can be distinguished.

It was shown earlier by Rao {\it et al.}\cite{durga2}, that the singlet state in not the least decohered state for all $\Delta$. One can find values of $\Delta$, where non-maximally entangled states have larger decoherence time scale in comparison to the maximally entangled states. Hence the natural question would be to know whether measurement on the partial entangled basis by Alice can improve the average fidelity given in Eq.\ref{avfid}. In the next section we shall take up the task of evaluating the average fidelity of teleportation when Alice measures her two-qubits in the non-maximally entangled basis.

\subsection{Measurement in partially entangled basis}
We shall consider the following one-parameter class of states as basis for Alice measurement, given by
\begin{eqnarray}
\hspace{-5cm}
|S^r_0\rangle&=& \frac{1}{\sqrt{1+r^2}}[|\uparrow\downarrow-r\downarrow\uparrow\rangle],~ |T^r_0\rangle = \frac{1}{\sqrt{1+r^2}}[r|\uparrow\downarrow+\downarrow\uparrow\rangle], \nonumber \\
|T^r_+\rangle&=&\frac{1}{\sqrt{1+r^2}}[r|\uparrow\uparrow+\downarrow\downarrow\rangle], ~
|T^r_-\rangle = \frac{1}{\sqrt{1+r^2}}[|\uparrow\uparrow-r\downarrow\downarrow\rangle]. \nonumber 
\end{eqnarray}
The density matrix for the above basis has a general representation
\begin{eqnarray}
\rho_{aA}={\hat{\mathcal{I}}\over 4} + {1\over 2}{P}^z_a(r){S}^z_a - {1\over 2}{P}^z_A(r){S}^z_A + \sum_{m=1}^{3}\Pi_{mm}(r)S^m_{a}S^m_{A}.
\end{eqnarray}
Note that the correlation matrix $\Pi$ is diagonal only for $r \in \mathbb{R}$. 

Similar to the analysis done in the earlier section, we find the average fidelity of the teleported state after the four measurements to be
\begin{eqnarray}
\mathcal{F}^{S^r_0}_{av}(t) &=& \frac{1}{2}+\frac{(1+r)^2+2r}{6(1+r^2)}\left[1-\frac{t^2}{\tau^2_0}(1-\Delta)\right],\nonumber \\
\mathcal{F}^{T^r_0}_{av}(t) &=& \frac{1}{2}+\frac{(1+r)^2+2r}{6(1+r^2)}\left[1-\frac{t^2}{\tau^2_0}\left(1+\frac{\Delta(1+r^2)}{(1+r)^2+2r}\right)\right],\nonumber \\
\mathcal{F}^{T^r_\pm}_{av}(t) &=& \frac{1}{2}+\frac{(1+r)^2+2r}{6(1+r^2)}\left[1-\frac{t^2}{\tau^2_0}\left(1+\frac{2\Delta r}{(1+r)^2+2r}\right)\right],\nonumber 
\end{eqnarray}
where $1/\tau^2_0 = \frac{1}{3}\langle \hat{I}^2_{\mathcal{E}} \rangle( K^2_a+K^2_A) $.
One can see that for $r=1$, the form of above equations reduce to that obtained in the earlier section. Because of the fraction $\frac{(1+r)^2+2r}{6(1+r^2)}$ appearing in the above expressions the average fidelity of teleportation in the case of partial entangled measurement is always less than the Bell-state measurement i.,e $\mathcal{F}^{S_0}_{av}(t) \ge \mathcal{F}^{S^r_0}_{av}(t)$ and similarly for all other Bell-states. In contrast to the earlier case where all the three Bell-states belonging to the triplet sector gave the same average fidelity, we find that they get further distinguished with $T^r_0$ giving one value of $\mathcal{F}_{av}(t)$ and $T^r_+, T^r_-$ different. This difference is because of the trace of the correlation matrix being different for these states for $r\ne 1$.

Though partially entangled states have long decoherence time-scales in comparison to maximally entangled states in some range of $\Delta$, teleportation always prefers Bell-basis measurement, if Alice and Bob initially share a maximally entangled state. The individual polarizations of qubits $P_a$ and $P_A$ though non-zero, did not contribute to $\mathcal{F}_{av}(t)$ because of the averaging performed on the surface of Bloch sphere.
\begin{figure}[htb]
\begin{center}
   \includegraphics[width=8.0cm]{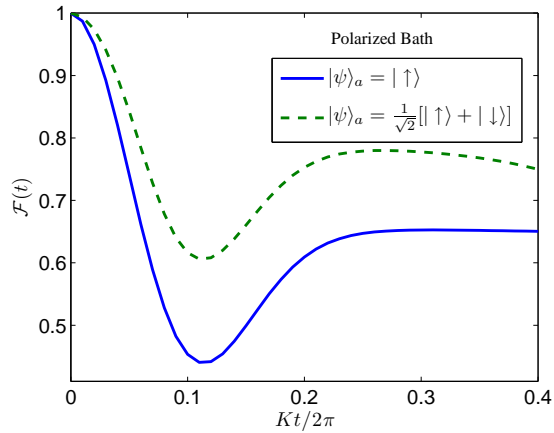}
\end{center}
 \caption{(Color online)Fidelity as a function of time. We have considered two different states which are to be teleported in the presence of a polarized bath. The initial state of the bath is $\rho_\mathcal{E} = \sum_I \lambda_I |I;I\rangle\langle I;I|$, where $\lambda_I \sim I^2\exp(-2I^2/N)$, and in the state $|I;I\rangle$ the first index corresponds to the spin value and the second value corresponds to the $\hat{z}$ component of the spin. The bath is composed of $N=22$ spins and $K^2 = K^2_a + K^2_A$. }
 \end{figure}
\subsection{Effect of initial polarizations on the fidelity}
In the earlier section we have considered the bath to be completely unpolarized, i.e., there is no preferred direction from the bath and hence all the unknown states with a given purity can be teleported with equal fidelity. In contrast if the bath is polarized there will be a preferred direction chosen by the bath because of which the fidelity of various state which Alice wishes to teleport can be different. A detailed analytical analysis for the polarized baths as done for the case of unpolarized baths is quite difficult. Instead, we have shown the effects of initial bath state on the short and long time behaviour of fidelity of different unknown states numerically. In Fig.2 we have plotted $\mathcal{F}(t)$ for two different unknown states, when the bath is polarized along $\hat{z}$ direction. It can be seen from Fig.2 that when the polarization of the unknown state is perpendicular to the polarization direction of the bath, both the decoherence time scale and the saturation value of $\mathcal{F}(t)$ are large in comparison to the situation where the polarization of the unknown state is parallel to the bath direction. In one case the fidelity is always great that the limiting value $2/3$, where as for the other case $\mathcal{F}(t)$ becomes smaller than $2/3$ in certain time regime but finally saturates close to the limiting value.

It was shown earlier by many authors (see for example \cite{loss}, \cite{durga1}) that the polarization of the bath helps reduce the qubit decoherence. In the case of quantum dots a bath polarization of $60\%$ has been achieved experimentally \cite{polexp}. It can be seen from Fig.2, that in the presence of polarized baths teleportation of all unknown states cannot be done under the same footing. Instead, one has to narrow down to some specialised states (which may belong to certain regions of the Bloch sphere) thereby reducing the unknowness of the state being teleported. If a third party (an eavesdropper) has control over the bath polarizations, and by knowing that the state has been teleported faithfully to Bob he can approximately estimate the unknown state. This can be quite harmful for the secure communication which the quantum teleportation promises.

In conclusion we have studied the average fidelity for teleporting an unknown state to Bob, when Alice's qubits see a common environment. Because of the common interaction with the bath, the average fidelity varies with the Bell state measurement performed by Alice. Even after Bob's operation, the state still has the information of Alice's measurement. This feature cannot be seen both in the decoherence free teleportation and teleportation through local noisy channels (separate baths). The singlet measurement always gives Bob the unknown state with high fidelity only when both the qubits of Alice interact either ferromagnetically or antiferromagnetically with bath. In contrast if one of the Alice's qubits interact ferromagnetically and the other anti-ferromagnetically then measurement of Bell states belonging to the triplet sector will give better fidelity. Instead of Bell measurement if Alice performs measurement on partially entangled basis, then the measurement further distinguishes the states, but the fidelity is smaller than obtained from Bell measurement. The common bath is responsible for an indirect interaction between the qubits of Alice. Eventhough not shown here there will be a generation of entanglement (bath induced entanglement) between the initially uncorrelated qubits of Alice, because of the effective exchange interaction between the qubits induced by the bath.

\begin{acknowledgments}
One of the authors Rao, would like to acknowledge the financial support provided by the J. C. Bose fellowship.
\end{acknowledgments}

.
 

\begin{thebibliography}{14}
\bibitem{book}{\it Theory of open quantum systems} by H. P. Breuer and F. Petruccione, (Oxford University Press, London  2002)
\bibitem{spbath} N V Prokof'ev and P C E Stamp, Rep. Prog. Phys. {\bf 63}, 669(2000). 
\bibitem{durga1}D. D. Bhaktavatsala Rao, V. Ravishankar, and V. Subrahmanyam,  Phys.
\text{\text{Re}}v. A {\bf 74}, 22301 (2006).
\bibitem{bruer}H. P. Breuer, D. Burgarth, and F. Petruccione, Phys. Rev. B 70, 045323 (2004).
\bibitem{bennet}C. Bennett {\it et al.}, Phys. Rev. Lett. {\bf 70}, 1895 (1993)
\bibitem{recent1}S. Oh, S. Lee and H. W. Lee, Phys. Rev. A {\bf 66}, 022316 (2002).
\bibitem{recent2}P. Badzia̧g, M. Horodecki, P. Horodecki and R. Horodecki  Phys. Rev. A {\bf 62}, 012311 (2000) 
\bibitem{nent}S. Muralidharan and P. K. Panigrahi, Phys. Rev. A {\bf 77}, 032321 (2008) 
\bibitem{qdottele1}K. W. Choo and L. C. Kwek, Phys. Rev. B {\bf 75}, 205321 (2007) 
\bibitem{qdottele2}O. Sauret, D. Feinberg and T. Martin, Eur. Phys. J. B {\bf 32}, 545(2003)
\bibitem{benati}F. Benatti, R. F. Floreanini, and M. Piani Phys. \text{\text{Re}}v. Lett {\bf 91}, 70402 (2003).
\bibitem{durga2}D. D. Bhaktavatsala Rao, V. Ravishankar, and V. Subrahmanyam, Phys. Rev. A {\bf 75}, 052338 (2007).
\bibitem{polexp}A. S. Bracker {\it et al.}, Phys. Rev. Lett. {\bf 94}, 047402 (2005).
\bibitem{loss}W. A. Coish and D. Loss, Phys. Rev. B {\bf 72}, 125337 (2005) 
 \end{thebibliography}
\end{document}